\documentclass[12pt]{iopart}
\usepackage{epsfig}
\usepackage{graphics}
\usepackage{amstext}

\begin{document}
\title[]{Overview of the BlockNormal Event Trigger Generator}

\author{J~W~C~McNabb, M~Ashley, L~S~Finn, E~Rotthoff, A~Stuver, T~Summerscales, P~Sutton\footnote[1]{current address: LIGO Laboratory, California Institute of Technology, Pasadena CA 91125, USA}, M~Tibbits, K~Thorne, K~Zaleski}

\address{Center for Gravitational Wave Physics, The Pennsylvania State University,University Park, PA 16802, USA}

\ead{\mailto{mcnabb@gravity.psu.edu}}

\begin{abstract}
        In the search for unmodeled gravitational wave bursts, there are a
variety of methods that have been proposed to generate candidate events from
time series data.  Block Normal is a method of identifying candidate events
by searching for places in the data stream where the characteristic statistics 
of the data change.  These change-points divide the data into blocks in which 
the characteristics of the block are stationary.  Blocks in which
these characteristics are inconsistent with the long term characteristic 
statistics are marked as Event-Triggers which can then be
investigated by a more computationally demanding multi-detector analysis.
\end{abstract}

\section{Finding Unmodeled Gravitational Waves}
The ongoing search for the direct detection of gravitational waves using
Earth based experiments involves the analysis of observations made at 
a variety of different detectors.  These observations are time series samples 
of the detector state that are then processed by various 
means to identify gravitational wave candidates.  Broadly speaking the searches
for gravitational waves may be broken up into two categories, those searches 
that are based upon a model, such as the search for gravitational waves from 
inspiraling binaries, and those searches that endeavor to find gravitational
waves without using a model.  The latter category of searches are often 
referred to as ``burst'' searches.  These searches typically seek to identify 
portions of data that are, for a short period, anomalously ``loud'' in
comparison to the surrounding data.

Identifying a gravitational wave burst in the absence of a source model is an 
involved and potentially computationally expensive process.  This is
especially true when the ratio of signal power to noise power is low.  A 
convenient and natural approach to mitigating the computational expense of 
identifying such bursts divides the problem of detection into two parts. 
In the first part, an inexpensive procedure is used to identify candidate 
sections of data that trigger the second part of the analysis.  The second 
part of the analysis focuses on the subintervals of data identified by the 
first.  It is a more computationally complex and expensive analysis that 
either discards the candidate or identifies it as a gravitational wave burst.  
In this way the first part of the analysis carried out by a so-called 
``event-trigger generator'', performs triage on the data that must be analyzed 
by the more complex second stage of the analysis.

\section{The BlockNormal Pipeline}
BlockNormal is an Event Trigger Generator that analyzes data in 
the time domain and searches for moments in time where the statistical 
character of the time series data changes.  In particular, BlockNormal 
characterizes the time series between change-points by the mean($\mu$) and 
variance($\nu$) of the samples.  

Change-points are thus demarcation points, separating ``blocks'' of data that 
are consistent with a distribution having a given mean and variance, which
differs from the mean and/or variance that best characterizes the data in an
adjacent block.  The onset of a signal in the data will, because it is 
uncorrelated with the detector noise, increase the variance of the time series 
for as long as the signal is present with significant power.  In this way 
blocks with variance greater than a ``background'' variance mark candidate 
gravitational wave bursts.

Since candidate gravitational wave bursts are identified with changes in the 
detector noise character it is best if the detector noise is itself stationary 
and white. The BlockNormal analysis thus starts by identifying long segments 
of data, epochs, which are relatively stationary.  This
process involves comparing adjacent stretches of data of fixed a duration long 
relative to the expected duration of a gravitational wave burst and asking 
whether adjacent stretches have consistent means and variances.  In order to 
avoid any bias that might come from analyzing outliers in the tails of the 
distribution (where one might expect any true signal to be located), the 
mean and variance are computed on only those samples that are within the 2.5th 
and 97.5th percentiles of the sample values.  If two consecutive stretches 
are inconsistent at the $95\%$ confidence level, the begining
of the first stretch is used to define a new stationary segment.

Segments thus defined are split into a set of frequency bands
whose lower band edge is heterodyned to zero frequency. This base-banding 
allows for a crude determination of the frequency of any identified triggers.  
Line and other spectral features are removed, either by Kalman filtering or by 
regression against diagnostic channels, and the final data whitened with a 
linear filter.

\subsection{Finding Change-Points}
Once the data has undergone the base-banding and whitening process, the search
for change-points begins in earnest.  The method employed is 
similar to that described in~\cite{smith} 
and relies on a Bayesian analysis of the relative 
probability of two different hypotheses:

\begin{itemize}
\item $M_1$, that the time series segment $X_N$ is drawn from a distribution 
characterized by a single mean and variance; and
\item $M_2$, that the time series segment $X_N$ consists of two continuous and 
adjacent subsegments each drawn from a distribution characterized by a 
different mean and/or variance.
\end{itemize}
Given the time series segment $X_N$, consisting of $N$ samples, we write the 
probability of hypothesis $M_1$ as $P(M_1|X_N)$ and the probability of 
hypothesis $M_2$ as $P(M_2|X_N)$. The odds of $M_2$ compared to $M_1$ is thus:
\begin{eqnarray}
  \rho_2 = \frac{P(M_2 | X_N)}{P(M_1 | X_N)}
\end{eqnarray}
Applying Bayes Theorem and simplifying this becomes:
\begin{eqnarray}
  \rho_2 = \frac{P(X_N|M_2)}{P(X_N|M_1)} \frac{P(M_2)}{P(M_1)} = 
  \frac{P(X_N|M_2)}{P(X_N|M_1)} \gamma_2
\end{eqnarray}
where, $\gamma_2=\frac{P(M_2)}{P(M_1)}$ is independent of the data $X_N$ 
itself,although it does depend on the number of samples $N$.

If $M_2$ is true, then $M_1$ will be a good hypothesis for two subsets of the 
data, one from sample 1 to $j-1$, denoted $X_{1,j-1}$ and another from 
$j$ to $N$, $X_{j,N}$.  Then we can write:
\begin{eqnarray}
  \rho_{2,j} = \gamma_2 \frac{P(X_j|M_1)P(X_{N-j} | M_1)}{P(X_N|M_1)} \\
    \rho_2=\sum_{j=1}^{N-1} \rho_{2,j}
\end{eqnarray}

To calculate $\rho_2$ we thus need only be able to calculate $P(X_N|M_1)$
for arbitrary time series $X_N$ The probability that a given data set is drawn 
from a normal distribution with unknown mean and variance is equal to:
\begin{eqnarray}
P(X_{N}|M_1)=\int d\sigma\int d\mu(2\pi\sigma^{2})^{-\frac{N}{2}}P(\mu,\sigma)\prod_{k=0}^{N-1}e^{-\frac{(x[k]-\mu)^{2}}{2\sigma^{2}}}
\end{eqnarray}

Where $P(\mu,\sigma)$ is the a priori probability that the mean takes on a value
$\mu$ and the variance a value $\sigma^2$.  With the usual uninformative priors
for $\mu$ and $\sigma$ ($P(\mu)\propto\alpha$ and $P(\sigma)\propto \beta\sigma^{-1}$) the integral for $P(X_N|M_1)$ can be evaluated in closed form:
\begin{eqnarray}
P(X_N|M_{1})=\frac{\alpha\beta}{\sqrt{N}}[2\pi N(\overline{x^{2}}-\overline{x}^{2})]^{-\frac{(N-1)}{2}}I_{N-2} \\
I_{N} \equiv (N-1)!! \left\{\begin{array}{ll}1 & \mbox{$N$ odd} \\ \sqrt{\frac{\pi}{2}} & \mbox{$N$ even} \end{array} \right.
\end{eqnarray}
where $\alpha \equiv P(\mu)$ an $\beta \equiv \sigma P(\sigma)$.

The value of $\rho_2$ is therefore the odds that there is a change-point
in $X_n$ to there not being any change-points, and the value $\rho_{2,j}$ is 
related to the odds that there is a change-point at sample $j$ to there not 
being one anywhere in $X_N$.  The calculated value of $\rho_2$ is compared to 
a threshold, $\rho_T$, and if greater than this threshold, a change-point is 
considered to be at the sample with the largest value of $\rho_{2,j}$.

\begin{figure}
  \centering
  \label{signal}
  \includegraphics[width=\textwidth]{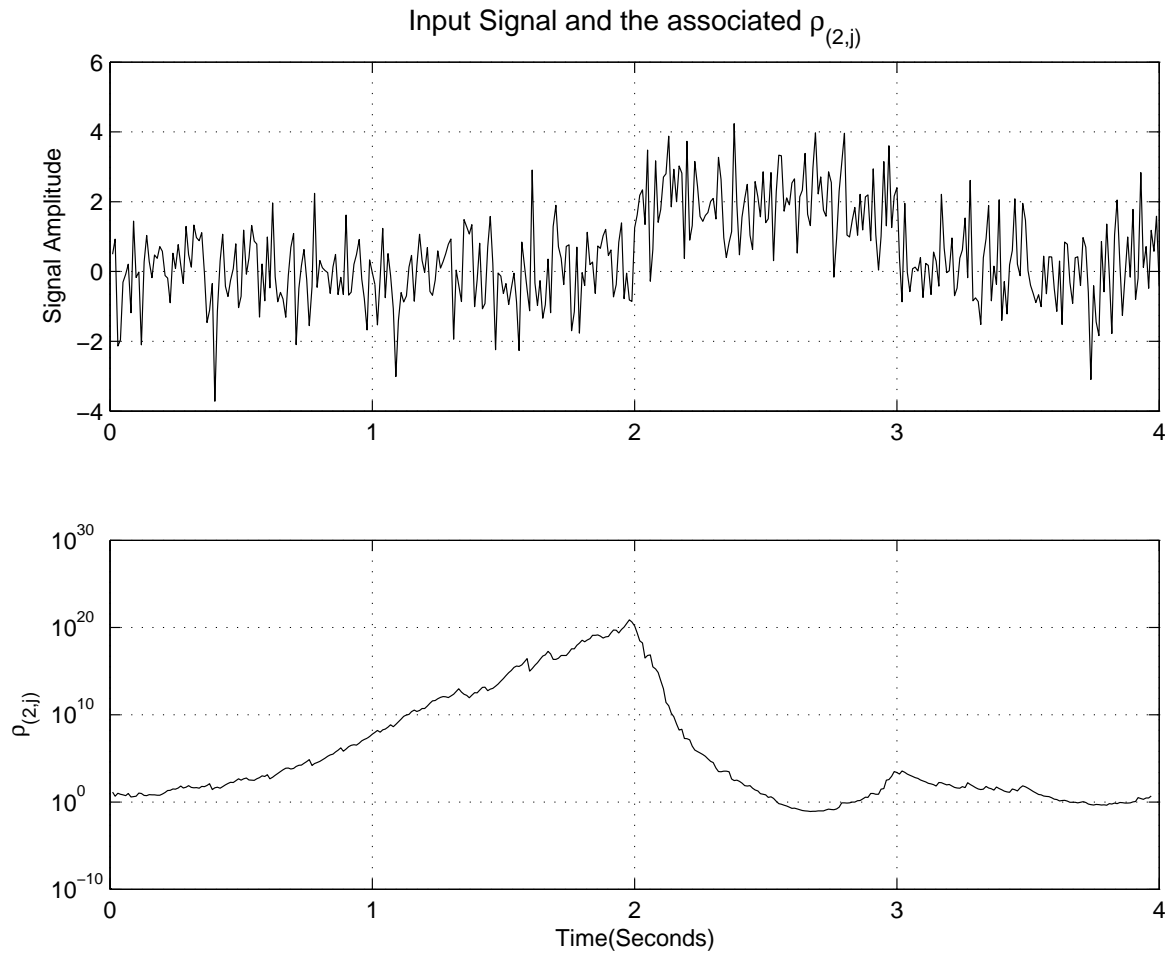}
%  \epsfbox{signal.eps}
  \caption{
(top) A sample of simulated data consisting of blocks of white noise 
with a mean of two between two and three seconds and zero elsewhere.  
(bottom) The associated figure of merit, $\rho_{2,j}$ as a function of the 
hypothetical change point time.}
\end{figure}

Figure~\ref{signal} shows some simulated data along with $\rho_{2,j}$ for
that data.  The two peaks in the value of $\rho_{2,j}$ corresponds 
to where the mean of the simulated noise changes.

This process either leaves $X_N$ free of change-points, or it divides the data
into two subsets.  In the latter case, the BlockNormal Pipeline repeats
the change-point analysis on these subsets and all subsequent subsets, until
either no more change-points are found, or the subset is less than 4 data 
points long.  By this iteration method BlockNormal breaks the data at these 
change-points into a set of blocks, each of which is free of any change point.
A final refinement step is taken where successive pairs of blocks are 
analyzed to check that the change-point would still be considered significant 
over the subset of the data that is contained within the two blocks.
   
\subsection{Blocks, Events, and Clusters}
BlockNormal identifies blocks --- time series segments between successive 
change-points --- are well characterized by a mean, a variance, a frequency 
band, a start time, and a duration. To identify unusual blocks, the means and 
variances are compared to the mean and variance, $\mu_0$ and $\sigma_0^2$ of 
all the data in their band from the epoch in which they were found.  
BlockNormal defines ``Events'' to be blocks in which the following condition 
holds true for a value, $C \equiv \max (\sigma^2, (\mu - \mu_0)^2 )$, that is 
used to characterize the block:
\begin{eqnarray}
  C > E_T \sigma_0^2
\end{eqnarray}
Here, $E_T$, is called the event threshold and is a free parameter in the
algorithm which adjusts the sensitivity in defining what ``unusual'' means.
There are a limited number of reasonable other possible threshold requirements 
based on the three characteristics of a block, $\mu$,$\sigma$,$D$, however, at 
this time these have not been explored.

Once events have been identified, immediately adjacent events in the same
frequency band are clustered together, with a peak-time for the cluster
defined by the central time of the block with the largest value of $C$.  
A cluster ``Energy'' is a sum over the $i$  blocks that 
comprise it:
\begin{eqnarray}
  \text{Energy} = \sum_{i} D_i \delta t \left[ (\mu_i^2-\mu_0^2) + \frac{D-1}{D}(\sigma_i^2 - \sigma_0^2)\right] 
\end{eqnarray}
where $D_i$ is the duration in samples.

\subsection{Defining Triggers}

\begin{figure}
  \centering
  \label{COINC}
  \includegraphics[width=\textwidth]{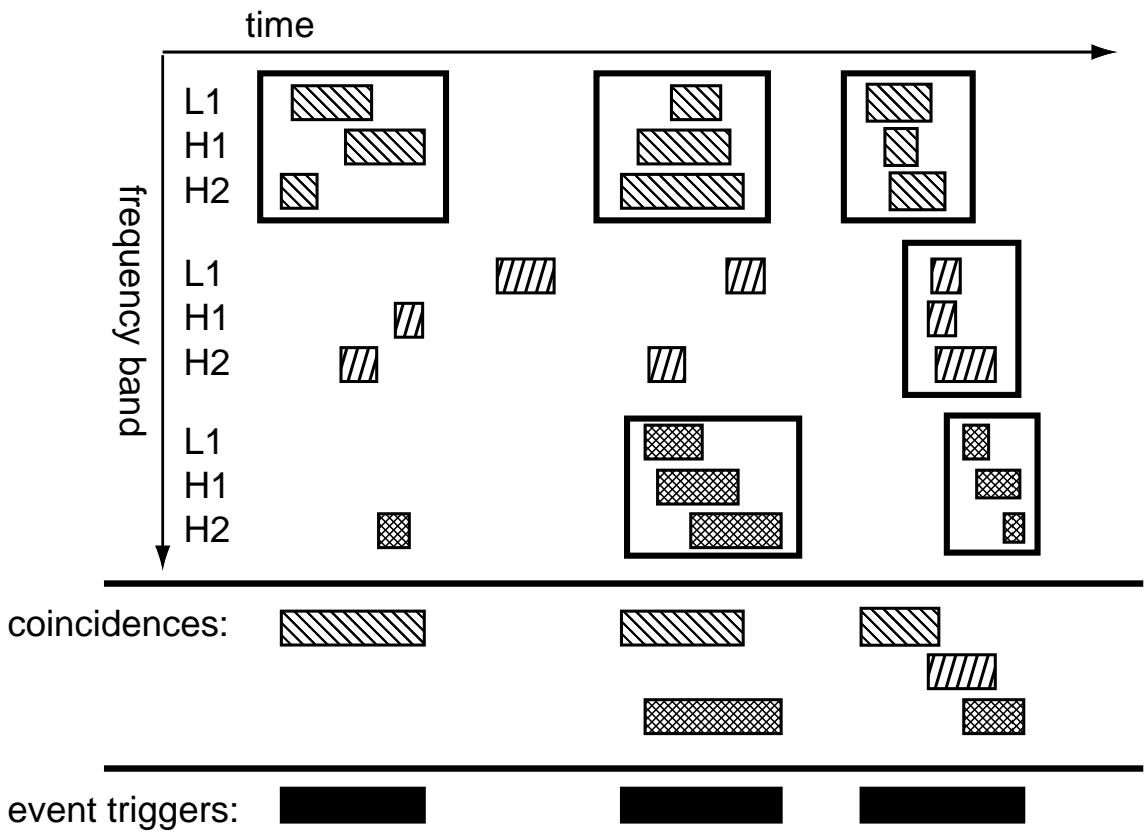}
%  \epsfbox{coincidence.eps}
  \caption{An illustration of how coincidence between different LIGO detectors 
and different frequency bands is used to create individual triggers.}
\end{figure}

A key factor in building confidence in any identification of gravitational 
waves is the presence of a signal in different detectors.  Using this 
``coincidence'' as the basis for further reducing the number of periods of 
interest, BlockNormal requires that there be coincidence in time between 
events in the same band but different detectors before a ``trigger'' is 
formed.  Triggers from different bands are merged if they overlap in time into 
a single trigger.  Figure~\ref{COINC} illustrates how this coincidence and 
merging step works using the three LIGO interferometers.

\section{Conclusion}
The BlockNormal event trigger generator is a time domain analysis in 
base-banded data that identifies blocks in time which are well characterized 
by a mean and variance. blocks with means and variances that are unusually 
large compared to the mean and variance of the much longer data segment 
containing them are marked for consideration as being unusual events.  
Several coincident events in different detectors together form a trigger which
can be used to define periods of interest that a more computationally intense 
analysis can use to reduce the total computing time needed to search for 
unmodeled gravitational wave events.

\ack
This work was supported by the Center for Gravitational Wave Physics,
the International Virtual Data Grid Laboratory, and the National
Science Foundation under award PHY 00-99559. The International Virtual
Data Grid Laboratory is supported by the National Science Foundation
under cooperative agreement PHY-0122557; the Center for Gravitational
Wave Physics is supported by the National Science Foundation under
cooperative agreement PHY 01- 14375.

\end{document}